# Multi-domain Integrative Swin Transformer network for Sparse-View Tomographic Reconstruction

Jiayi Pan[1], Heye Zhang[1], Weifei Wu[2], Zhifan Gao[1] and Weiwen Wu[1,3,*]

[1]School of Biomedical Engineering, Sun Yat-sen University, Shenzhen, Guangdong, China
[2]Department of Orthopedics, the people's Hospital of China Three Gorges University, the First People's Hospital of Yichang, Hubei, China
[3]Lead Contact
*Correspondence: wuweiw7@mail.sysu.edu.cn

**Summary:** Decreasing projection views to lower X-ray radiation dose usually leads to severe streak artifacts. To improve image quality from sparse-view data, a Multi-domain Integrative Swin Transformer network (MIST-net) was developed in this article. First, MIST-net incorporated lavish domain features from data, residual-data, image, and residual-image using flexible network architectures, where residual-data and residual-image sub-network was considered as data consistency module to eliminate interpolation and reconstruction errors. Second, a trainable edge enhancement filter was incorporated to detect and protect image edges. Third, a high-quality reconstruction Swin transformer (i.e., Recformer) was designed to capture image global features. The experiment results on numerical and real cardiac clinical datasets with 48-views demonstrated that our proposed MIST-net provided better image quality with more small features and sharp edges than other competitors.

## I. INTRODUCTION

Computed Tomography (CT) has been widely used in medical diagnosis and industrial detection fields because of its excellent imaging ability [1]. Especially in 2020, CT became an essential technology to detect and diagnose the COVID-19 [2]. Although CT scans provide practical and accurate diagnostic results, they are also increasingly harmful to human bodies with radiation dose [3]. An effective approach to reduce radiation dose is sparse-view CT reconstruction [4,5], where only part of projection data is used for image reconstruction. In this case, traditional reconstruction algorithms such as filtered back-projection (FBP) [6] leads to serious streaking artifacts as well as low image quality.

Since the emergence of artificial intelligence [7-9], many deep learning-based methods have been developed to improve the quality of sparse-view CT reconstruction [10-12]. They can be divided into three categories: image domain restoration [13], dual-domain restoration [14,15], and iterative reconstruction methods [16,17]. For image domain restoration methods, they are also called post-processing methods. They directly process low quality images as the input and ground truth as output, which means that this kind of methods don't need raw projections data. The typical reconstruction networks include FBPconvNet [13], Densenet Deconvolution Network

(DD-Net) [11] and residual encoder-decoder convolutional neural network (RED-CNN) [10]. Besides, Zhang et al. [18] used a generative adversarial network (GAN) to remove sparse views artifacts. Wang et al. [19] presented a limited-angle CT image reconstruction algorithm based on a U-net convolutional neural network, which can effectively eliminate noise and artifacts while preserving image structures. Since there is no projection data playing the game, it is difficult to accurately recover image details and features with streak artifacts removal.

The dual-domain deep reconstruction methods usually concentrate to reconstruct high-quality images by considering both projection and image domains simultaneously. For example, Hu et al. [20] proposed a Hybrid Domain neural Network (HDNet), which recovered projection and image information successively. Liu et al. [21] presented a lightweight structure utilizing spatial correlation. Zhang et al. [22] designed a hybrid-domain convolutional neural network for limited-angle computed tomography. Wu et al. [23] presented a Dual-domain Residual-based Optimization NEtwork (DRONE), which performed well in edge preservation and details recovery. The dual-domain network can be also applied to 3D reconstruction [24]. However, the final images may suffer from secondary artifacts due to the introduced errors from projections interpolation.

Inspired by the classic iteration reconstruction, deep reconstruction networks can also be designed as unfolding deep iterative reconstruction. Cheng et al. [25] accelerated iterative reconstruction with the help of deep learning. Chen et al. [26] presented a Learned Experts' Assessment-based Reconstruction Network (LEARN) for sparse-view data reconstruction. Zhang et al. [27] extended the LEARN model to a dual-domain version, named LEARN++. Xiang et al. [28] proposed a Fast Iterative Shrinkage Thresholding Algorithm (FISTA) for inverse imaging problems. Unrolling iterative reconstruction methods can suppress noise and artifacts to improve image quality. Nevertheless, iterative reconstruction methods need huge GPU memory, which leads to them being difficult to 3D geometry.

In this work, we proposed a Multi-domain Integrative Swin Transformer network (MIST-net) to reconstruct high quality CT images from sparse-view projections. The overall of our network architecture consists of three key components: initial recovery, data consistency correction and high-fidelity reconstruction. In the initial recovery, a data-extension encoder-decoder block is first employed to extend sparse-view data to full-view projection data by deep interpolation. Then, an end-to-end edge enhancement reconstruction sub-network reconstructed the initial image with sparse artifacts removal and image edges preservation. However, projection domain interpolation may introduce errors to introduce unexpected artifacts. Therefore, the data consistency module, which consists of two residual sub-networks (one for residual projection estimation, the other for residual image correction), was introduced to reduce errors and improve structural details. Although CNN-based deep learning reconstruction methods have provided good performance, it cannot learn global and long-range image information interaction well due to the locality of convolution operator [29,30]. Fortunately, transformers [31] have such ability for modeling long-range information and show good performance in nature language processing (NLP) tasks [31-33]. The proposal of Vision Transformer [34] shows that transformers can take the place of convolutions in some image processing tasks [35-37]. Therefore, we introduce a Hierarchical Vision Transformer using Shifted Windows (Swin) [37] in the high fidelity reconstruction module to capture long-range dependencies.

Compared with developed CNN-based deep networks for the sparse-view reconstruction in the past few years, our MIST-net is innovative in several aspects. First, an encode-decode [38] structure is employed in the initial recovery module to extract deep features in both data and image domains simultaneously. Specifically, an edge enhancement reconstruction network in the image domain was designed to recover the image edge. Second, both data-residual and image-residual networks are used in the data consistency module to eliminate errors in both projection and image domain, which contribute to artifacts reduction and subtle structure recovery. Third, the Swin reconstruction transformer (Recformer) extracts both shallow and deep features in the image domain to ensure final reconstruction results of our MIST-net.

The organization of the paper is as follows. In section II, the main reconstruction results from our MIST and competitors are reported. We also implement the detailed ablation study including numerical and real cardiac data to show the advantages of our MIST-net. We further do noise analysis experiments to verify the robustness of the model. In section III, we discuss our results and make conclusions of this work. In the **EXPERIMENTAL PROCEDURES** section, we introduce basic theories and then describe our proposed MIST-net carefully.

## II. RESULTS

In this study, we developed a Multi-domain Integrative Swin Transformer network (MIST-net) to reconstruct CT images from ultra-sparse view projections. To obtain initial recovery images, two CNN-based sub-networks in both projection domain and image domain were designed. Then, we designed a dual-domain residual network to eliminate errors and noise. Our proposed Swin reconstruction transformer (i.e., Recformer) sub-network refined intermediate results. Figure 1 demonstrates the overall architecture of our MIST-net, more details can be found in the **EXPERIMENTAL PROCEDURES**.

Our model was designed and trained in Python using the PyTorch framework. All experiments were run on a PC with 48G NVIDIA RTX A6000 GPU, Intel(R) Xeon(R) Gold 6242R CPU @ 3.10GHz and 128GB RAM. The configuration of the training network is as follows. Our network was trained by Adam optimizer and the learning rate was set to 0.00025. The number of epochs was 50 and batch size was 1. FBPconvNet [13], HDNet [20], DDNet [11], FISTA [28] and LEARN [26], are treated as comparisons. The root mean square error (RMSE), peak signal-to-noise ratio (PSNR) and structure similarity index (SSIM) are introduced to quantitatively assess reconstruction results. Our code is publicly released on https://zenodo.org/record/6368099.

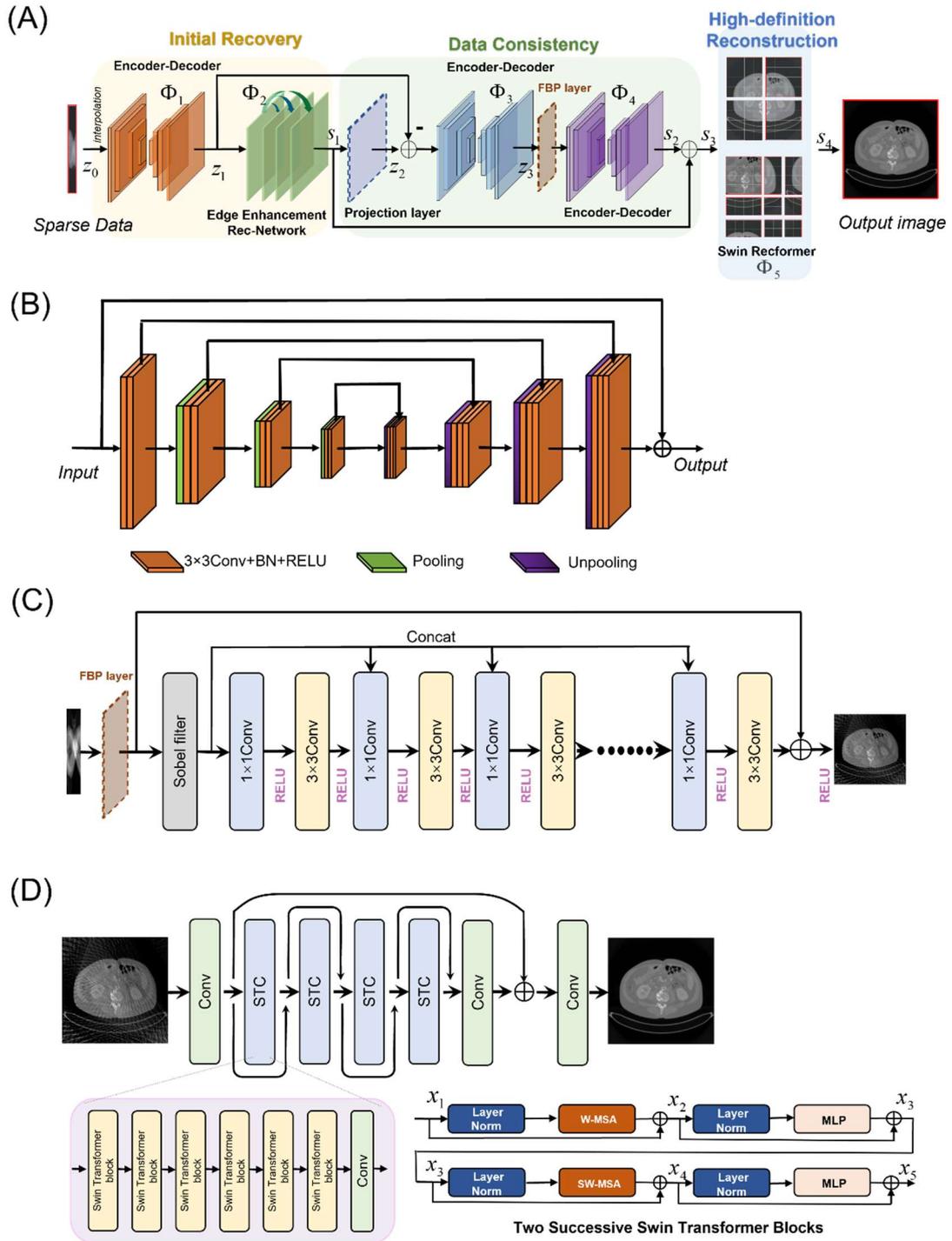

**Figure 1. The overall architecture of our proposed MIST-net.** (a) The pipeline of the proposed MIST-net. Our network has three modules: initial recovery, data consistency and high-definition reconstruction; (b) the encoder-decoder block with similar U-net architecture; (c) the edge enhancement reconstruction sub-network for recovering an initially-reconstructed CT image. (d) represents Swin Recformer network to reconstruct a high-quality image.

*A. Simulated Data Result*

To validate the feasibility of our proposed network, we train and test our model on 2016 NIH-AAPM-Mayo Low-dose CT Grand Challenge datasets [39]. The datasets come from Siemens Somatom Definition CT scanners at 120kVp and 200mAs. To generalize our model to real datasets, we rearranged the datasets with all scanning parameters and configuration being consistent with following real cardiac CT datasets. Specifically, the distances from x-ray source to the system isocenter and detector are set as 53.85cm and 103.68cm. The number of detector units and views are set to 880 and 2200 respectively. The size of reconstructed CT images is 512×512. Finally, a total number of 4,665 sinograms of 2,200×880 pixels were acquired from 10 patients at the normal dose setting, where 4,274 sinograms of 8 patients were employed for network training, and the rest 391 sinograms from other 2 patients for network testing. Our operation to obtain sparse data is as follows: for every sinogram of 2,200×880 pixels, we sample every 30 views until 48×880 pixels have been collected. We also sample every 10 views to obtain a projection of 144×880 pixels, which is used as a projection label.

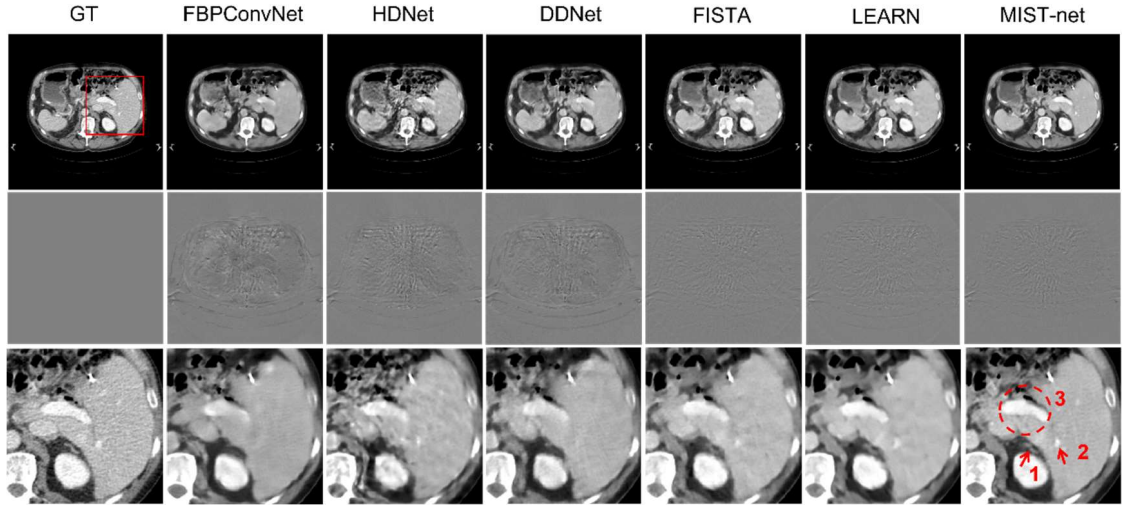

**Figure 2. Visualizations of sparse-view reconstruction in case 1 by using different methods.** The 1st-7th columns represent ground truth (GT), FBPconvNet, HDNet, DDNet, FISTA, LEARN and MIST-net counterparts from 48 views. The 2nd row shows the difference images relative to the GT and 3rd represents the extracted region-of-interest (ROI) from 1st row images. The display windows for the reconstructed and difference images are [-160 240] HU and [-90 90] HU.

Figures 2 and 3 demonstrated the representative reconstruction results with cases #1 and #2 from patients #1 and #2 of different reconstruction networks. It was clearly observed that FBPConvNet removed most of the artifacts caused by sparse-views, but the image boundaries and details were further destroyed. HDNet obtained better images but it leads to excessive image smoothing. DDNet was able to recover some image details, but its results still contained a few unacceptable artifacts. Unlike above-mentioned methods, FISTA and LEARN as two of advanced unrolled deep reconstruction methods have better performance in sparse-view reconstruction since it can effectively improve image quality with richer details and clearer edges. However, some tiny features were still lost. In contrast to competitors, our MIST-net improved image quality with the best details and edges.

To display the advantages of the MIST-net, the region of interests (ROIs) was extracted and magnified in Figure 2 and Figure 3. Firstly, one can find that the magnified structures marked by arrows "1" and "2" were badly blurred and destroyed by FBPConvNet, HDNet and DDNet in Figure 2. In contrast to FBPconvNet, HDNet and DDNet, FISTA and LEARN achieved better images. However, the results of FISTA-net and LEARN were still inferior to our MIST network. Besides, the FBPconvNet, HDNet and DDNet missed details and over-smoothen the edge of tissues in the circle "3". Compared with FBPconvNet, HDNet, and DDNet, FISTA-net and LEARN almost eliminated artifacts and achieved a better image. However, it was still found that the structure of tissue was slightly fuzzy, which exposed its weakness on edge recovery. On the contrary, our MIST-net obtained the best reconstructed result with clear edges and rich details in the image region indicated by the circle "3".

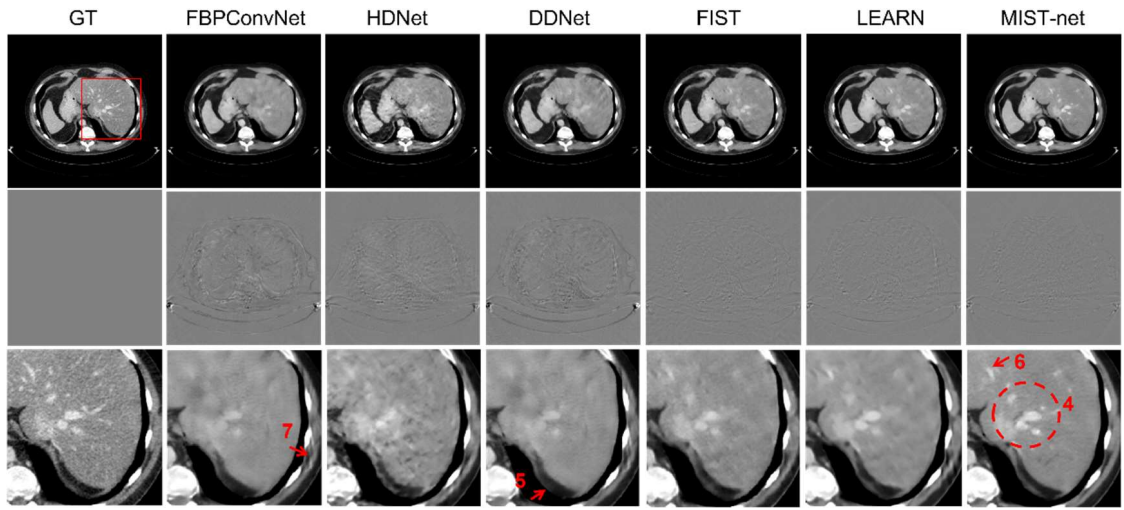

**Figure 3. Visualizations of sparse-view reconstruction in case 2 by using different methods.** The 1st-7th columns represent Ground Truth, FBPconvNet, HDNet, DDNet, FISTA, LEARN and MIST-net counterparts from 48 views. The 2nd row shows the difference images relative to the GT and the 3rd row shows the magnification ROIs. The display windows for the reconstructed and difference images are [-160 240] HU and [-90 90] HU.

On the other hand, in Figure 3, the image structure indicated by circle "4" could demonstrate the advantages of MIST-net in terms of structural fidelity. The image feature highlighted by circle "4" was almost lost or damaged by FBPconvNet, HDNet and DDNet. FISTA and LEARN could retain a few structural features but they still looked blurry. In addition, the feature with the arrow "5" showed that DDNet and FBPconvNet produced a grey intensity shift, while FISTA, LEARN and MIST-net can address this case. To further explain the superiority of our MIST-net in terms of fine texture retain, the easily overlooked details marked with arrows "6", "7" were emphasized in Figure 3. With these highlight structures, it can be inferred that FBPconvNet, HDNet and DDNet smeared small image features. FISTA and LEARN recovered some image features missed by above-mentioned methods but the finer image features such as that indicated by the arrow "6" were destroyed. As a result, compared with other methods, MIST-net achieved the best results that look quite similar to the ground truth. More reconstruction experiment results can be found in **Appendix A of Supplemental Experimental Procedures (see figure S1 and figure S2)**.

Table 1

Quantitative evaluation of 48 projections reconstruction results from two simulated cases

| Views | | | FBPconvNet | HDNet | DDNet | FISTA | LEARN | MIST-net |
|---|---|---|---|---|---|---|---|---|
| 48 | RMSE↓ | case1 | 27.1508 | 24.1323 | 24.6031 | 18.1993 | 18.0470 | **16.2775** |
| | | case2 | 28.7080 | 22.4121 | 29.1107 | 17.7193 | 17.5947 | **15.8242** |
| | PSNR↑ | case1 | 38.3953 | 39.4190 | 39.2512 | 41.8699 | 41.9429 | **42.8392** |
| | | case2 | 37.0957 | 39.2461 | 36.9747 | 41.2868 | 41.8392 | **42.2693** |
| | SSIM↑ | case1 | 0.9573 | 0.9625 | 0.9602 | 0.9744 | 0.9760 | **0.9800** |
| | | case2 | 0.9647 | 0.9672 | 0.9635 | 0.9752 | 0.9784 | **0.9818** |

We also made a quantitative evaluation of all methods, and the results were quantified in Table 1. It was observed that our MIST network produced the best results than the FBPconvNet, HDNet, DDNet, FISTA and LEARN methods. Table 1 demonstrated that FBPconvNet and DDNet achieved the worst performance in PSNR, SSIM, and RMSE, which indicated that the performance of post-processing methods was affected severely by sparse views artifacts. Compared to FBPconvNet only concentrating on image domain, HDNet stacked two U-Net structures respectively in both projection domain and image domain. HDNet had better scores than FBPconvNet, which benefited from the effectiveness of the hybrid domain processing. Meanwhile, FISTA and LEARN certainly outperformed FBPconvNet, HDNet and DDNet in all evaluations because of iterative processing. In Table 1, our proposed MIST-net method has the smallest RMSEs and the biggest PSNRs and SSIMs than those competitors. These quantitative results validated the advantages of the proposed MIST-net demonstrating the best performance. More statistical quantitative results from all simulation testing datasets were given in Table 2 and they demonstrated our MIST-net can obtain best performance. Finally, we also done the experiments with more views (i.e., 64 views) and their results were given in **Appendix A** and **B of Supplemental Experimental Procedures**.

Table 2

Quantitative evaluation of 48 projections reconstruction results from simulated testing datasets

| Views | Methods | RMSE | PSNR | SSIM |
|---|---|---|---|---|
| 48 | FBPconvNet | $27.5520 \pm 3.8158$ | $38.0544 \pm 1.4500$ | $0.9596 \pm 0.0095$ |
| | HDNet | $23.9574 \pm 3.3687$ | $39.2675 \pm 1.2331$ | $0.9649 \pm 0.0076$ |
| | DDNet | $25.9597 \pm 3.7597$ | $38.5814 \pm 1.6092$ | $0.9611 \pm 0.0091$ |
| | FISTA | $19.4109 \pm 2.1991$ | $41.0691 \pm 1.0308$ | $0.9730 \pm 0.0046$ |
| | LEARN | $17.7858 \pm 2.0657$ | $41.8307 \pm 1.0600$ | $0.9782 \pm 0.0043$ |
| | MIST-net | $\mathbf{16.1408 \pm 1.7620}$ | $\mathbf{42.6700 \pm 1.0895}$ | $\mathbf{0.9817 \pm 0.0037}$ |

*B. Clinical Cardiac Validation*

To further verify the performance of MIST-net, the real dataset used in [40] was used. The curved cylindrical detector contains 880 units. There are 2200 views as the full scan. The diameter of field-of-view (FOV) covers 49.8 × 49.8 $cm^2$ with an image matrix of 512 × 512 pixels. The distance from the X-ray source to the system isocenter and the detector array were 53.85 cm and 103.68 cm. Since we have trained the reconstruction network using AAPM datasets, here, we transferred the trained network to evaluate the reconstruction performance using real dataset, which can benefit to evaluate the generalization ability of our model. We also extracted 48 views from the clinical short scan to test our MIST-net for sparse-view CT imaging. It is worth mentioning that the data distribution of real datasets is different from simulated datasets. Thus, we pre-processed the clinical cardiac dataset to keep its data distribution consistent with training datasets. Specifically, we first normalize the clinical dataset. The obtained data is then mapped to the numerical distribution of the simulated data.

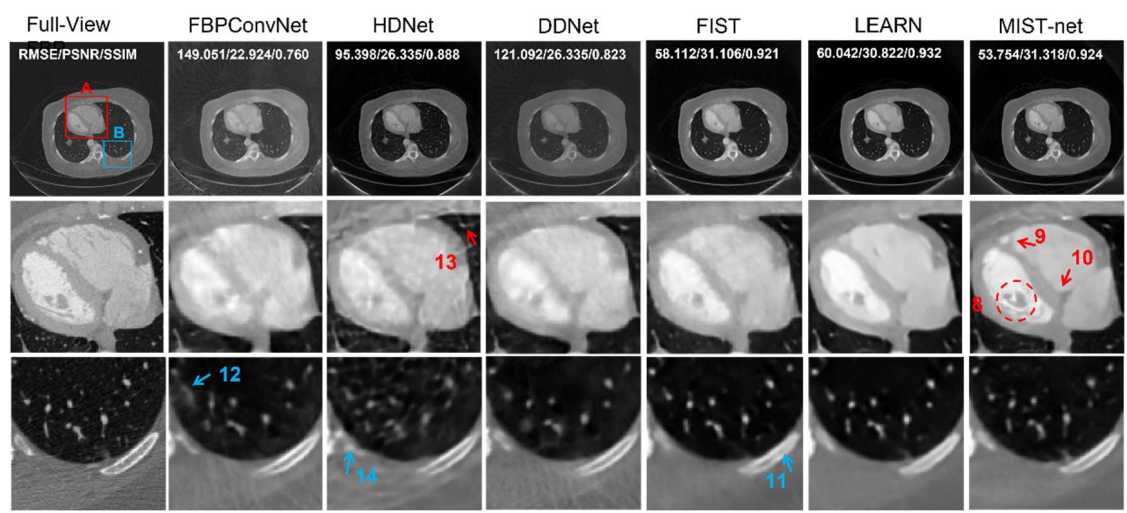

**Figure 4. Clinical cardiac CT reconstructions from sparse-view data by using different networks.** The 1st-7th columns stand for the FBP reconstruction from full-view data, FBPconvNet, HDNet, DDNet, FISTA, LEARN and MIST-net counterparts from 48 views. The second and third rows show ROIs. The display windows for the reconstructed images are [-800 1000] HU.

Figure 4 showed the outputs from 48 views using different reconstruction methods. The full-view FBP reconstruction also contained some noise and short-scan artifacts [41]. Compared with these competitors, our proposed MIST-net achieved the best reconstruction results. Firstly, as shown in the circle "8", MIST-net achieved the structure closest to the ground truth, which adequately embodies the advantages of MIST-net in terms of structural fidelity. In addition, the image feature marked by the arrow "9" was failed to be recovered by FBPconvNet, HDNet, DDNet, FISTA and LEARN, however, the proposed MIST recovered the structure, which showed that MIST-net was good at details and feature recovery. Again, MIST-net reconstructed clearer image edges and showed finer structures. The clearest edges indicated by the arrow "10", which cannot be obviously reconstructed by other competitors, strongly confirmed the advantages of the proposed MIST-net. On the contrary, the edges in the reconstructed images were hard to distinguish from FBPconvNet, HDNet and DDNet. FISTA and LEARN provided a better reconstruction in the arrow "10" but the details were still compromised. Furthermore, the feature indicated by the arrow "11" was destroyed from FISTA-net.

Compared with our proposed model, other non-iterative methods performed unsatisfyingly. Obviously, FBPconvNet could not recover the image structure indicated by the arrow "12". Benefited by a dual domain design, HDNet suppressed most of artifacts and noise but it still caused fuzziness as well as sparse-view artifacts, which were clearly indicated by the arrows "13" and "14". What's more, benefited from the iterative mechanism, the reconstructed image from FISTA and LEARN were better than that obtained by FBPconvNet, HDNet and DDNet. In edge restoration, FISTA and LEARN were still worse than our proposed network. The real experiments further demonstrated that the proposed network performed consistently better than other competitors in practice. It also demonstrated the power of the transformer in the sparse-view CT reconstruction.

### C. Ablation Exploration and Generalization Analysis

For analyzing and benchmarking the proposed network MIST-net. Here, we further focused on the ablation explorations to validate effectiveness of different modules. First, the DU-RecNet represents a simplified MIST-net which uses two encoder-decoder blocks to replace the edge enhancement reconstruction network and Swin Recformer network. The MU-RecNet denoted the modified DU-RecNet by removing the data consistency module only. All networks were trained in the same way. As shown in Figure 5, the residual domain processing is helpful for artifact reduction and detail recovery. Regarding the results indicated by circle "1" and arrow "2", DU-RecNet could recover better features than MU-RecNet, since residual domain sub-network was designed to eliminate interpolation errors. The RMSE, PSNR and SSIM results further explained the advantage of residual domain sub-network.

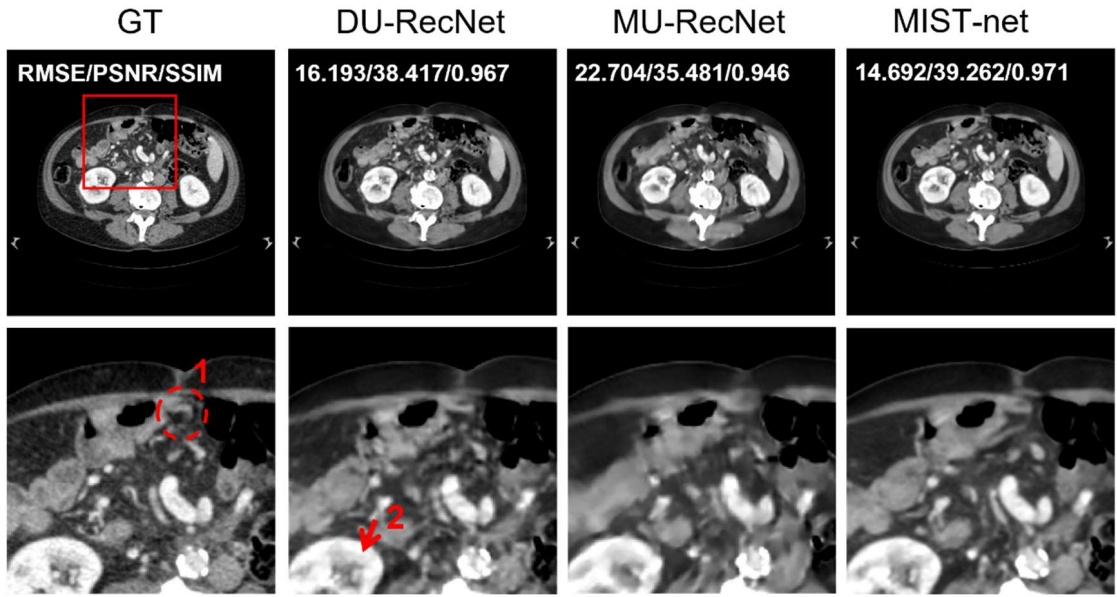

**Figure 5. Reconstruction results in Case 3.** The 1$^{st}$-3$^{rd}$ columns represent Ground Truth, reconstructions from DU-RecNet and MU-RecNet counterparts with 48 views. The 2$^{nd}$ row shows the ROIs. The display window for the reconstructed images is [-160 240] HU.

We also constructed an EE-RecNet to verify the effectiveness of the edge enhancement reconstruction network. Compared to DU-RecNet, EE-RecNet only added the edge enhancement reconstruction network. The architecture of EE-RecNet was similar to MIST-net except that the final Swin Recformer network was replaced by an Unet. Figure 6 showed reconstruction results from DU-RecNet, EE-RecNet and MIST-net. All models were trained in the same manner. As shown in the top of Figure 6 (case #4), the result from DU-RecNet contained a few artifacts due to sparse-view down sampling. The edge enhancement reconstruction module indeed reduces artifacts in the image domain and helps to overcome edge over-smoothness. Additionally, we found that the image region marked with circle "3" was destroyed by DU-RecNet. Both EE-RecNet and MIST-net reconstructed the general outline, but EE-RecNet lost some details by the arrow "4". Furthermore, from bottom of Figure 6 (case #5), one observed that Swin transformer was also important for high-contrast structural recovery, low-contrast feature reconstruction and textural details preservation. The detail indicated by arrow "5" was blurred by DU-RecNet while MIST-net could reconstruct it well. In addition, features marked by the arrows "6" and "7" were very similar to the ground truth, but they were blurry in the images reconstructed by DU-RecNet and EE-RecNet.

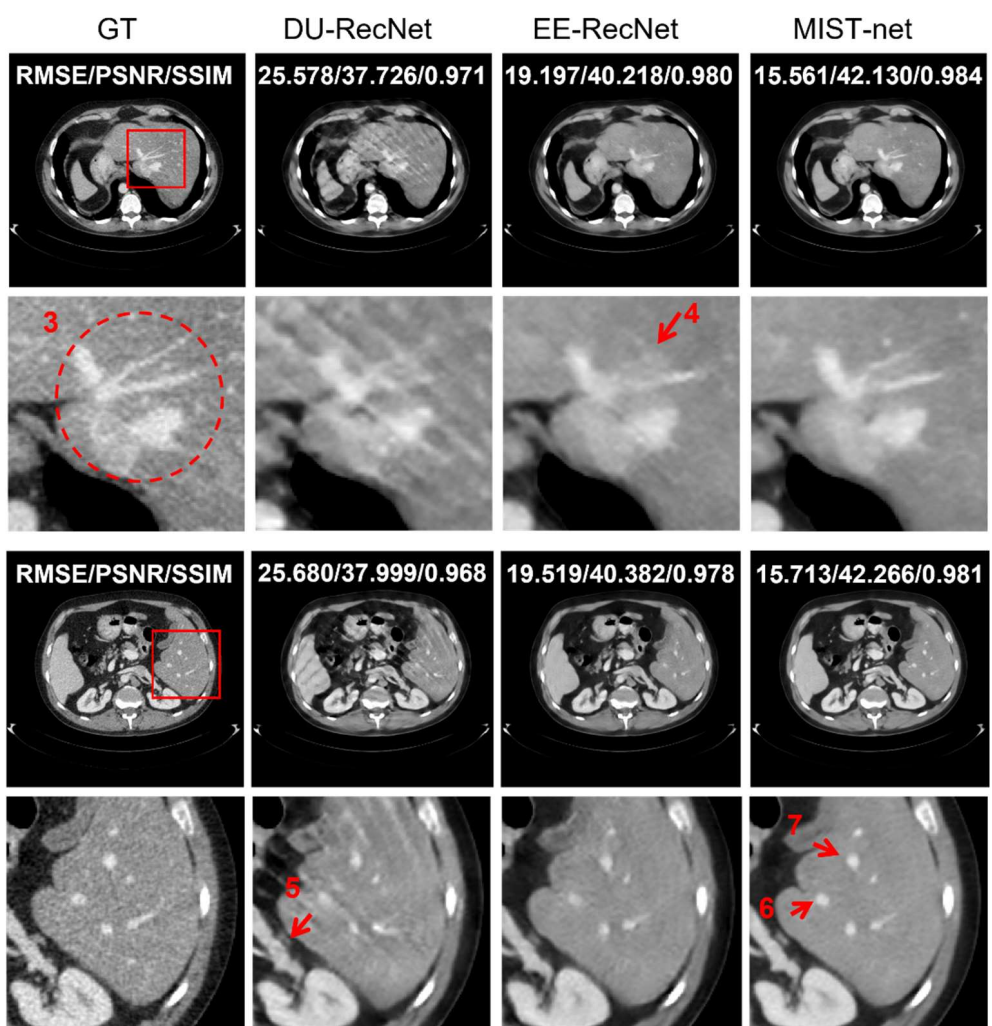

**Figure 6. Comparison of different networks reconstruction results in Case 4 (top) and 5 (bottom).** The 1st-4th columns represent Ground Truth, reconstructions from DU-RecNet, EE-RecNet and MIST-net counterparts from 48 views. The second row of each part shows the ROIs. The display window for the reconstructed images is [-160 240] HU.

The RMSE, PSNR and SSIM metrics were computed as well to confirm the gain with the edge enhancement reconstruction network. The quantitative results in terms of RMSE, PSNR and SSIM have clearly illustrated the merits of our proposed MIST-net.

The statistical quantitative evaluations of ablation experiments were computed in terms of RMSE, PSNR and SSIM, and their results were summarized in Table 3. It can be seen that our MIST-net can obtain the best quantitative statistical results in terms of mean and standard deviation than other networks.

Table 3

Quantitative evaluation of ablation experiments

| Views | Network | RMSE | PSNR | SSIM |
|---|---|---|---|---|
| 48 | DU-RecNet | $19.3569 \pm 3.9817$ | $41.2178 \pm 2.1800$ | $0.9773 \pm 0.0064$ |
|  | MU-RecNet | $25.4315 \pm 3.1333$ | $38.7374 \pm 1.3734$ | $0.9651 \pm 0.0066$ |
|  | EE-RecNet | $18.0518 \pm 1.9057$ | $41.6966 \pm 1.1847$ | $0.9789 \pm 0.0042$ |
|  | MIST-net | $\mathbf{16.1408 \pm 1.7620}$ | $\mathbf{42.6700 \pm 1.0895}$ | $\mathbf{0.9817 \pm 0.0037}$ |

To demonstrate the influence of different modules, we also did an ablation experiment on the real cardiac CT dataset. Figure 7 showed the clinical cardiac reconstructed images from 48 views using relative methods. The performance of MU-RecNet was compromised, and the edges and details were hard to distinguish. DU-RecNet reconstructed observed features but still caused hazy edges, which were clearly indicated by an arrow in Figure 7. EE-RecNet recovered details but produced sparse-view artifacts. Compared with competitors, MIST-net delivered the best image quality and evaluation indicators.

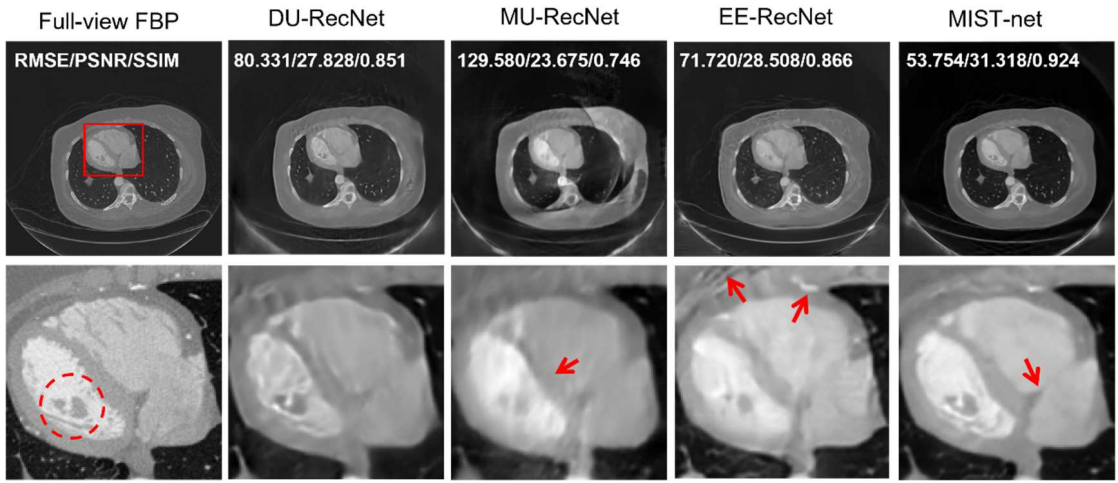

**Figure 7. Comparison of different networks reconstruction results in the real dataset.** The 1st-5th columns stand for the FBP reconstruction from full-view data, DU-RecNet, MU-RecNet, EE-RecNet and MIST-net counterparts from 48 views. The second row shows ROIs. The display windows for the reconstructed images are [-800 1000] HU.

To demonstrate the advantages of our MIST-net with similar parameters and memory, we design a pure CNN structure as a baseline, which uses five ResUnets in all parts of the procedure. Then, we study the effect of edge enhance sub-network and dual transformers with similar parameters. The result can be found in Table 4. The "Baseline+Edge-Ehance" is similar to overall structure of EE-RecNet expect the number of parameters. Compared to MIST-net, "Baseline+Dual-SwinRec" uses Swin Transformer in both edge enhancement sub-network and the last reconstruction block. With similar parameters, Table 4 shows that both edge enhancement Rec-Network and Swin Transformer play important roles in controlling image quality.

Table 4
Comparison between the ablation candidates with similar size

| Networks | RMSE | PSNR | SSIM | #Params. |
|---|---|---|---|---|
| Baseline | 19.7047 ± 3.1251 | 40.9957 ± 1.7905 | 0.9757 ± 0.0061 | 12.2M |
| +Edge-Enhance | 17.6826 ± 1.9234 | 41.8785 ± 1.2036 | 0.9795 ± 0.0042 | 12.3M |
| +Dual-SwinRec | 18.0210 ± 2.3201 | 41.7362 ± 1.5209 | 0.9789 ± 0.0051 | 12.4M |
| MIST-net | **16.1408 ± 1.7620** | **42.6700 ± 1.0895** | **0.9817 ± 0.0037** | 12.0M |

To further verify the effect of Swin-Recformer as the last module. We have designed Swin-Recformer modules of different complexity and introduced MIST-Tiny, MIST-Small, MIST-Base and MIST-Large. The design of their initial recovery and data consistency blocks is exactly the same, and the difference is only in the Swin Transformer module. In this paper, we use MIST-B as the proposed MIST-net. The architecture hyper-parameters of these model variants are:

- MIST-T: C=96, layer numbers= {2,2,6,2}, head numbers= {3,6,12,24}
- MIST-S: C=96, layer numbers= {2,2,18,2}, head numbers= {3,6,12,24}
- MIST-B: C=96, layer numbers= {6,6,6,6}, head numbers= {6,6,6,6}
- MIST-L: C=96, layer numbers= {8,8,8,8}, head numbers= {8,8,8,8}

where C is channel number of hidden layers in the first stage. The quantitative evaluation and number of parameters are listed in Table 5. The experimental results show that MIST-B achieved the best results. In addition, we observed that larger model (MIST-L) leads to worse results, which possibly due to overfitting.

Table 5
Comparison between Swin Rec-former modules of different computational complexity

| Views | Network | RMSE | PSNR | SSIM | #Param. |
|---|---|---|---|---|---|
| 48 | MIST-T | 16.9113 ± 1.8407 | 42.2660 ± 1.1869 | 0.9804 ± 0.0040 | 10.5M |
|  | MIST-S | 16.4780 ± 1.7965 | 42.4916 ± 1.1954 | 0.9814 ± 0.0038 | 12.1M |
|  | MIST-B | **16.1408 ± 1.7620** | **42.6700 ± 1.0895** | **0.9817 ± 0.0037** | 12.0M |
|  | MIST-L | 16.8268 ± 1.7739 | 42.3063 ± 1.1594 | 0.9807 ± 0.0039 | 13.0M |

We also study residual data in the data consistency module. As shown in Figure 1(A), i.e. in our network structure, the residual data $z_1 - z_2$ is inputted to the encoder-decoder block $\Phi_3$. As a comparison, we use $z_0 - z_2$ as an input and make $\Phi_3$ an interpolation network. The quantitative analysis can be found in Table 6. The results show that $z_1 - z_2$ works better as residual data inputted to $\Phi_3$.

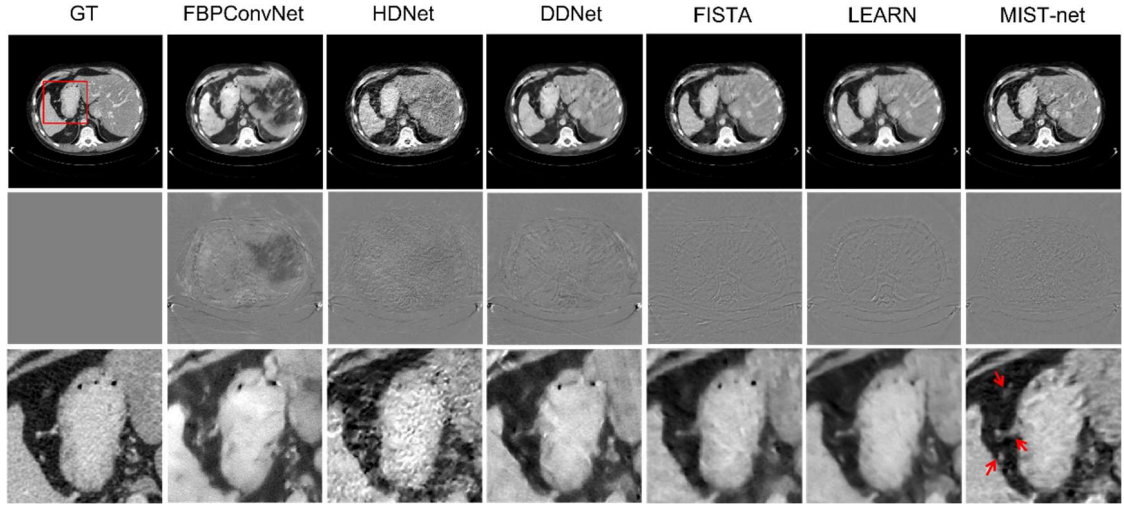

**Figure 8. The generalization of different deep reconstruction networks against noise on simulation datasets.** The 1st-7th columns stand for the ground truth, FBPconvNet, HDNet, DDNet, FISTA, LEARN and MIST-net counterparts from 48 views. 1st-3rd rows represent reconstructed results, difference images, and the magnified ROIs.

Table 6
Comparison between different designs of residual data

| Residual Data | RMSE | PSNR | SSIM |
|---|---|---|---|
| $z_0 - z_2$ | $26.9013 \pm 4.1702$ | $38.2671 \pm 1.4324$ | $0.9634 \pm 0.0153$ |
| $z_1 - z_2$ | $\mathbf{16.1408 \pm 1.7620}$ | $\mathbf{42.6700 \pm 1.0895}$ | $\mathbf{0.9817 \pm 0.0037}$ |

The generalization ability is also an important issue for deep learning-based image reconstruction in practice. In this study, the Gaussian noise was added to the images, where the mean and variance are set as 0 and 0.01. Then, the noisy images were employed to verify the ability of reconstruction models against noise attacks during the testing process. Figure 8 showed the reconstructed results with different networks. The structures marked with the arrows show that details were blurred by noise using FBPconvNet, HDNet, DDNet, FISTA and LEARN. Our proposed MIST-net can obtain better image quality than other competitors. The more detailed noise experiment results were given in **Appendix C of Supplemental Experimental Procedures (see figure S3)**.

## III. DISCUSSION

Deep learning has attracted rapidly increasing attention in the field of medical image analysis. Since 2016, convolutional neural network (CNN) based deep learning techniques have been extensively developed for tomographic imaging with sparse data, some of which were already approved by commercial scanners and translated into clinical practice. While CNN-based deep learning is impressive, it cannot learn global and long-range image information interaction well due to the locality of convolution operation (e.g. 3×3 or 5×5 region), which further results in that this kind of methods fail to capture global structures and features of tomographic imaging. Fortunately, the vision transformer can convert an image into a sequence to enhance capability of long-range modeling, which is also one of the inspirations for this paper.

This study presented a Multi-domain Integrative Swin Transformer network (MIST-net) for sparse-view CT reconstruction. Our primary contribution is that we first presented a Multi-domain Integrative Swin Transformer network (MIST-net) and then it is employed to sparse-view tomographic reconstruction. Our MIST-net reconstructs tomographic images with sparse-data, where both projection and image domains are respectively responsible for repairing projection data and restoring images in the initial sub-network reconstruction stage. Then, both residual-projection and residual-image domains sub-networks are used to eliminate measurement errors and make data be consistent. To retain features and enhance the edge of the tomographic image, an edge enhancement sub-network is introduced to avoid over-smoothness and edge blurring. More importantly, we proposed a Recformer (a novelty transformer) sub-network to capture global features and structures of tomographic imaging. Our work first demonstrated the feasibility of transformer-based tomographic imaging with sparse data as well extinguished reconstruction performance. The results showed that the proposed network could effectively reduce streaking artifacts caused by sparse-view projection, and recover image features and details.

We verified our approach with both simulated and clinical datasets, showing that it outperforms CNN-based methods such as FBPconvNet, DDNet, HDNet, FISTA and LEARN. FBPconvNet and DDNet represent the performance of image post-processing methods. HDNet as a dual-domain based deep reconstruction method encoded projection domain and image domain information simultaneously. FISTA and LEARN are unrolled iterative deep learning and they provide state-of-the-art reconstruction results. Compared with competitors, our MIST-net achieved the best quantitative performance. We also compare the complexity and runtime of all competitors (Table S2). Our network runs faster than LEARN and DDNet. Compared with HDNet, the advantages are also obvious because HDNet needs to train two networks separately, which makes the training process more complicated. Compared with FBPConvNet, since our network is larger, the comparison doesn't seem fair. To address this issue, we designed a MIST-Tiny (10.4M) in subsequent ablation experiments. The MIST-Tiny has a similar size with FBPconvNet (9.8M) and HDNet (9.8M×2), and still show relatively good performance. Compared with FISTA and LEARN, as they are iterative methods, the low number of parameters is a major feature of them. However, multiple projection and back-projection operations cause a huge computational footprint, which makes it difficult to apply to clinical practice. So, we can't rely on network parameters to judge who is good or bad between our network and iterative methods. In ablation experiments, we have verified the effectiveness of proposed modules under the condition of consistent parameters and we performed ablation studies on the swin module alone. It is worth noting that the performance of MIST-Large is degraded compared to standard MIST-Base. The possible reason is that the larger model leads to overfitting. Finally, Gaussian noise experiments further exhibited better generalization ability of MIST-net over competitors.

Our work first demonstrated the feasibility of transformer-based tomographic imaging with sparse data as well extinguished reconstruction performance. In future, the transformer can be introduced into unrolled iterative reconstruction. As mentioned above, the computational cost of unrolled methods depends on the number of iterations (i.e. projection and back-projection operations). Thus, the easiest way to reduce the amount of computation is to reduce the number of iterations. However, for a simple CNN, it is difficult to train well with a small number of iterations. By introducing the transformer, the reconstruction network can capture long-range information and become easier to train than pure CNN. For example, the iterative method of CNN and transformer may get good results with fewer iterations (e.g. 5 or 7 iterations).

Although the proposed network has demonstrated good performance in sparse-view tomographic reconstruction, there are still some issues that need to be addressed. First, the proposed method needs more computational and memory cost than pure CNN-based methods, which become a challenge of transformed-based application in medical imaging. Second, transformers had strict requirements on input image size because of position encoding, which limits the flexibility of transformer-based deep reconstruction. Therefore, we did not use Swin Transformers in our projection domain because the size of sparse-view projection is always different. In addition, the transformer requires large datasets to show the unique advantages. However, medical image datasets with labels are scarce. Extending our method to iterative reconstruction and exploring the feasibility of our MIST-net with self-supervised strategy [42] for limited-angle CT [43] and low dose CT [44] is our future plans. In summary, this paper presents a MIST-net reconstruction model and it will encourage transformer-based applications in medical imaging.

## IV. EXPERIMENTAL PROCEDURES

### Resource availability
#### Lead Contact
Weiwen Wu, PhD (email:wuweiw7@mail.sysu.edu.cn).
#### Materials availability
The library in this study is made publicly available via Zenodo. (https://zenodo.org/record/6368099)
#### Data and Code Availability
All original code has been publicly released on https://zenodo.org/record/6368099. This paper analyses existing public data. These data are available at: https://www.aapm.org/GrandChallenge/LowDoseCT/.

## Methodology

### A. CT Imaging Model
The ideal mathematical model of CT imaging can be expressed as a discrete linear system:

$$\mathbf{y} = \mathbf{A}\mathbf{x} + \mathbf{b} \qquad (Equation\ 1)$$

where $\mathbf{x}$ represents a reconstructed CT image, and it can be expressed as $\mathbf{x} = [x_1, x_2, x_3, \ldots x_P]^T$. $\mathbf{y}$ stands for the measured projection data, and it can be written as $\mathbf{y} = [y_1, y_2, y_3, \ldots y_Q]^T$. $\mathbf{b}$ stands for projection noise, and $\mathbf{A}$ is a CT system matrix, which contains

Q×P elements. Due to the noise in y, the solution of Equation 1 can be obtained by minimizing the following objective function:

$$\min_{\mathbf{x}} ||\mathbf{Ax} - \mathbf{y}||_F^2 \qquad (Equation\ 2)$$

where $||\cdot||_F^2$ stands for the Frobenius norm. The ART or SART are usually employed to minimize the Equation 2 [45]. However, solving x directly from a very sparse projection data y is an under determined inverse problem, which may lead to the poor reconstructed image quality with streak artifacts. So, in the traditional analytical methods, a regularization term representing prior knowledge is usually introduced to obtain better reconstruction result, then we have

$$\min_{\mathbf{x}} \left( ||\mathbf{Ax} - \mathbf{y}||_F^2 + \beta_h h(\mathbf{x}) \right) \qquad (Equation\ 3)$$

There are two components in Equation 3, i.e. the fidelity term $||\mathbf{Ax} - \mathbf{y}||_F^2$ and the regulation prior knowledge term $h(\mathbf{x})$. The hyperparameter $\beta_h$ is designed to balance these two components. By the way, different reconstruction methods correspond different regularization priors such as total variation [46,47] and dictionary learning [43,48,49].

**B. Multi-domain Integrative Network**

In the projection domain, the main work is to complement and restore the sparse-view sinograms. For example, Dong et al. [50] completed the missing data with U-Net architecture, then a typical network further refined the raw image reconstructed from completed projections. The benefit of the projection domain deep neural network is that it can reduce the data error from the view of detector measurement. However, the interpolated projection with deep neural networks may introduce wrong measurement and further result in false positive and false negative diagnosis results. Unfortunately, the false results are also difficult to correct even if by a high-fidelity post-processing image domain network. To overcome this challenge, the residual-data domain sub-network is first considered to correct the data inconsistency of the initial reconstructed image. Indeed, the stage is beneficial to correct original data errors to overcome the data inconsistency. Furthermore, one residual-image domain sub-network further improves the reconstruction performance. Finally, Swin Transformer architecture can deeply characterize various latent features of the reconstructed image to capture local and global information of image-self.

**C. Edge Detection Operator**

The edge of images was one of the most important features, including a wealth of internal information especially in medical images, for example, the edge of tumors helps to diagnose if it is benign or malignant. In this work, we introduced a Sobel filter [51] to overcome the problem of excessive edge smoothing. Sobel operator belonged to orthogonal gradient operator and its gradient corresponded to the first derivative. For a continuous function $g(a, b)$, where (a, b) indicates the position point, the gradient can further be expressed as a vector:

$$\nabla g(a,b) = \{G_a, G_b\} = \left\{\frac{\partial g}{\partial a}, \frac{\partial g}{\partial b}\right\} = \frac{\partial g}{\partial a}i + \frac{\partial g}{\partial b}j \quad (Equation\ 4)$$

$$\mathrm{mag}(\nabla g) = |\nabla g| = \sqrt{\left(\frac{\partial g}{\partial a}\right)^2 + \left(\frac{\partial g}{\partial b}\right)^2} \quad (Equation\ 5)$$

$$\theta(a,b) = \arctan\left(\frac{G_a}{G_b}\right) \quad (Equation\ 6)$$

where the $\mathrm{mag}(\nabla g)$ and $\theta(a,b)$ stand for the magnitude and direction angle of $\nabla g(a,b)$. The partial derivatives needed to be calculated for each pixel point by using the equation (6). Original Sobel operator contains a 3×3 vertical filter and a 3×3 horizontal filter, which has a maximum response to the vertical edge and the level edge respectively.

Liang et al. [52] proposed an edge enhancement-based densely connected network (EDCNN) and achieved good performance in low-dose CT denoising. Sharifrazi et al. [53] applied Sobel filters to achieve accurate detection of COVID-19 patients from CT images. Compared to other edge operators, the mechanism of Sobel is the differential of two rows or two columns, and it can fully enhance elements on both sides, which makes the edge seem more obvious. In this work, in addition to the level and vertical filter, we further add a diagonal filter to our network in [52].

**D. Vision Transformers**

The Transformer was first proposed for natural language processing. Transformer was similar to an encoder-decoder structure which consists of multi-head self-attention blocks, normalization layers and point-wise feed-forward networks. The Vision Transformer (VIT), proposed by Dosovitskiy et al. [34], which can be considered as the first vision transformer backbone for image classification. VIT had demonstrated the effectiveness of Transformer in CV tasks, although it required huge parameters and memory because the global computation led to quadratic complexity. To reduce the usage of GPU memory and number of calculation parameters, Swin Transformer [37] computed self-attention within local windows. The computational complexity of a global MSA module and a window based on image patches with the size of n × n are respectively recorded as $\Omega(4n^2C^2 + 2n^4C)$ and $\Omega(4n^2C^2 + 2M^2n^2C)$, where $n^2$, C and M are the number of pixels in an image patch, channel number of the hidden layers and window size, respectively. To solve the global modeling problem caused by local windows, Swin Transformer designed a shifted window to strengthen the connection between adjacent windows. Due to its impressive performance, the transformer has also been introduced to medical image processing. Chen et al. [54] proposed TransUNet, which claimed to be the first transformer based medical image segmentation Network. Recently, Eformer [55] used the self-attention and depth-wise convolution for better local context capture in medical image denoising. Our novel network MIST-net was developed to explore transformers in sparse-view data reconstruction. Again, we designed a Swin Recformer sub-network by combining the Swin Transformer and convolution layer to make full use of both shallow and deep features.

# MIST-net

Figure 1 illustrated the flowchart of our proposed MIST-net. There are three key components,

i.e., initial recovery, data consistency and high-definition reconstruction. Both initial recovery and data consistency have two sub-networks, one works in radon domain and the other was designed within images domain. Then we will describe each sub-network and more network details can be found in **Appendix D of Supplemental Experimental Procedures (see table S3, table S4 and table S5)**.

*1) Initial Recovery Module:* The architecture of this module is shown in Figure 1. The first block of this module is an encoder-decoder block with a linear interpolation at the beginning. This part is designed to restore a spare-view projection data $z_0$ to a full-view projection data $z_1$. This is achieved by a projection domain sub-network $z_1 = \Phi_1(z_0)$. At the first layer of initial reconstruction is the network-based image reconstruction, i.e., FBP layer. An edge enhancement reconstruction sub-network was employed to process the FBP layer output and further obtain a clearer and faithful image s₁. The initial recovery module can be expressed as follows:

$$s_1 = \Phi_2(\Phi_1(z_0)) \qquad (Equation\ 7)$$

where $\Phi_2$ represents the reconstruction sub-network from the initial recovery module.

*2) Data Consistency Module:* The errors are always introduced in projections and images because the interpolation in the Radon domain cannot accurately predict the missing original data. In addition, the following reconstruction sub-network in the images domain may result in false positive and negative results. These errors may cause the secondary artifacts to compromise the quality of images. Data consistency module consists of two parts: the projection-residual processing sub-network $\Phi_3$ and the image-residual processing sub-network $\Phi_4$. Here, we use two encoder-decoder blocks to handle residual data. The re-sampled residual data from $s_1$ can be expressed as $z_1 - z_2$, where $z_2$ represents the projection data from the image s₁. The estimated projection data residual $z_3$ of the data consistency module is expressed as:

$$z_3 = \Phi_3(z_1 - z_2) \qquad (Equation\ 8)$$

In addition to the data difference computed by the trained data residual network according to Equation 5, there is also an image difference between the output of the initial recovery module and desired image. Here, the image residual processing sub-network was further employed to reduce the data inconsistency. The output of the image residual processing sub-network $\Phi_4$ is as follow:

$$s_2 = \Phi_4(z_3) \qquad (Equation\ 9)$$

Then, we can get the middle reconstruction output by adding the output of initial recovery module to the results of data consistency module, we have that

$$s_3 = s_1 + s_2 \qquad (Equation\ 10)$$

*3) High-definition Reconstruction Module:* The sparse-view projection may lead to serious

streaking artifacts in CT images especially when the number of views is extremely scanty. The function of this module is similar to post-processing methods which learn mappings from poor images to clear images. In traditional image super-resolution and denoising tasks, SwinIR [56] has achieved great success by using the Swin Transformer as a backbone. In this part, we propose a hybrid architecture called Swin Recformer, which is based on both convolution layers and Swin Transformer layers to implement the image reconstruction task. As shown in Figure 1 (D), the Swin Recformer contains convolutional layers, Swin Transformer mixed convolution (STC) units and a few residual connections. Each STC unit consists of six transformer layers as well as one convolutional layer.

In this section, we will provide architecture details of the STC unit. A STC unit contains six Swin Transformer layers and a 3×3 convolution layer. The Swin Transformer block contains 2 core designs which are described below. First, Swin Transformer first designed a non-overlapping window-based multi-head self-attention (W-MSA) block, which can learn the long-range information correlation in a small-size window region (for example, a 16×16 feature map). Second, the shifted window-based multi-head self-attention (SW-MSA) block, which adds shifted windows to improve interactions between different windows. In the lower right of the Figure 1 (D), two successive Swin Transformer blocks are presented. Each Swin Transformer block is successively composed of LN layer, multi-head self-attention (MSA) mechanism, residual connection and MLP. The W-MSA mechanism and the SW-MSA mechanism make up two adjacent transformer blocks. With the shifted window partitioning design, consecutive Swin Transformer blocks are computed as:

$$\hat{x}^l = W - MSA(LN(x^{l-1})) + x^{l-1} \quad (Equation\ 11)$$
$$x^l = MLP(LN(\hat{x}^l)) + x^l \quad (Equation\ 12)$$
$$\hat{x}^{l+1} = SW - MSA(LN(x^l)) + x^l \quad (Equation\ 13)$$
$$x^{l+1} = MLP(LN(\hat{x}^{l+1})) + \hat{x}^{l+1} \quad (Equation\ 14)$$

where $\hat{x}^l$ and $x^l$ denote the output features of the W-MSA/SW-MSA module and the MLP module for block l, respectively; W-MSA and SW-MSA denote window based multi-head self-attention using regular and shifted window partitioning configurations, respectively. Same as the traditional transformer methods, self-attention is computed as follows:

$$\text{Attention}(Q, K, V) = \text{SoftMax}\left(\frac{QK^T}{\sqrt{d}} + B\right) \cdot V \quad (Equation\ 15)$$

where $Q, K, V \in R^{M^2 \times d}$ are the query, key and value matrices; d is the query/key dimension, and $M^2$ is the number of patches in a window. Since the relative position along each axis lies in the range [-M+1, M-1], we parameterize a smaller-sized bias matrix $\hat{B} \in R^{(2M-1) \times (2M+1)}$, and values in B are taken from $\hat{B}$. After six Swin Transformer blocks, a 3×3 convolutional layer was added to enhance the feature. Between adjacent STC units, a residual connection was used to aggregate feature maps generated from transformer and convolution.

*4) Loss Functions:* For sparse-view CT reconstruction, we optimize the parameters of MIST-net by minimizing the dual-domain mean square error (MSE). It can be written as follow:

$$\mathcal{L} = \|I_L - I_O\|_2^2 + \gamma\|P_L - P_O\|_2^2 \qquad (Equation\ 16)$$

where $\|I_L - I_O\|_2^2$ is the image MSE term. $I_L$ and $I_O$ stand for the label and output images, respectively. $\|P_L - P_O\|_2^2$ is the projection MSE, where $P_L$ and $P_O$ represent projection label and output. $\gamma$ is a weighting factor, here it was set as 0.1.

## V. ACKNOWLEDGMENTS

This work was partially supported by Shenzhen Science and Technology Program (Grant No. GXWD20201231165807008, 20200825113400001), Natural Science Foundation of Guangdong Province, China (2022A1515011384), National Natural Science Foundation of China (62101606，81702198).

## VI. AUTHOR CONTRIBUTIONS

J.P. proposed reconstruction method, debugged codes, carried out experiments and wrote draft manuscript; H.Z. provided benefit suggestions for the revision; WF.W. and Z.G commented on the paper and provided valuable suggestions; WW.W. initialized and supervised this project, debugged codes, and further revised the draft.

## VII. DECLARATION OF INTERESTS

The authors declare no competing interests.

## VIII. REFERENCES

1. Bakator, M., and Radosav, D. (2018). Deep learning and medical diagnosis: A review of literature. Multimodal Technologies and Interaction *2*, 47.
2. Long, C., Xu, H., Shen, Q., Zhang, X., Fan, B., Wang, C., Zeng, B., Li, Z., Li, X., and Li, H. (2020). Diagnosis of the Coronavirus disease (COVID-19): rRT-PCR or CT? European journal of radiology *126*, 108961.
3. Brenner, D.J., and Hall, E.J. (2007). Computed tomography—an increasing source of radiation exposure. New England journal of medicine *357*, 2277-2284.
4. Bian, J., Siewerdsen, J.H., Han, X., Sidky, E.Y., Prince, J.L., Pelizzari, C.A., and Pan, X. (2010). Evaluation of sparse-view reconstruction from flat-panel-detector cone-beam CT. Physics in Medicine & Biology *55*, 6575.
5. Bian, J., Wang, J., Han, X., Sidky, E.Y., Shao, L., and Pan, X. (2012). Optimization-based image reconstruction from sparse-view data in offset-detector CBCT. Physics in Medicine & Biology *58*, 205.
6. Katsevich, A. (2002). Theoretically exact filtered backprojection-type inversion algorithm for spiral CT. SIAM Journal on Applied Mathematics *62*, 2012-2026.
7. Litjens, G., Kooi, T., Bejnordi, B.E., Setio, A.A.A., Ciompi, F., Ghafoorian, M., Van Der Laak, J.A., Van Ginneken, B., and Sánchez, C.I. (2017). A survey on deep learning in medical image analysis. Medical image analysis *42*, 60-88.
8. Niu, C., Zhang, J., Wang, G., and Liang, J. (2020). Gatcluster: Self-supervised gaussian-attention network for image clustering. (Springer), pp. 735-751.
9. Niu, C., Cong, W., Fan, F.-L., Shan, H., Li, M., Liang, J., and Wang, G. (2021).

Low-dimensional manifold constrained disentanglement network for metal artifact reduction. IEEE Transactions on Radiation and Plasma Medical Sciences.
10. Chen, H., Zhang, Y., Kalra, M.K., Lin, F., Chen, Y., Liao, P., Zhou, J., and Wang, G. (2017). Low-dose CT with a residual encoder-decoder convolutional neural network. IEEE transactions on medical imaging *36*, 2524-2535.
11. Zhang, Z., Liang, X., Dong, X., Xie, Y., and Cao, G. (2018). A Sparse-View CT Reconstruction Method Based on Combination of DenseNet and Deconvolution. IEEE Trans Med Imaging *37*, 1407-1417. 10.1109/TMI.2018.2823338.
12. Wang, G., Ye, J.C., and De Man, B. (2020). Deep learning for tomographic image reconstruction. Nature Machine Intelligence *2*, 737-748.
13. Kyong Hwan, J., McCann, M.T., Froustey, E., and Unser, M. (2017). Deep Convolutional Neural Network for Inverse Problems in Imaging. IEEE Trans Image Process *26*, 4509-4522. 10.1109/TIP.2017.2713099.
14. Bertram, M., Rose, G., Schafer, D., Wiegert, J., and Aach, T. (2004). Directional interpolation of sparsely sampled cone-beam CT sinogram data. (IEEE), pp. 928-931.
15. Liu, J., Ma, J., Zhang, Y., Chen, Y., Yang, J., Shu, H., Luo, L., Coatrieux, G., Yang, W., and Feng, Q. (2017). Discriminative feature representation to improve projection data inconsistency for low dose CT imaging. IEEE transactions on medical imaging *36*, 2499-2509.
16. Yu, W., Wang, C., and Huang, M. (2017). Edge-preserving reconstruction from sparse projections of limited-angle computed tomography using ℓ 0-regularized gradient prior. Review of Scientific Instruments *88*, 043703.
17. Humphries, T., Winn, J., and Faridani, A. (2017). Superiorized algorithm for reconstruction of CT images from sparse-view and limited-angle polyenergetic data. Physics in Medicine & Biology *62*, 6762.
18. Xie, S., Xu, H., and Li, H. (2019). Artifact removal using GAN network for limited-angle CT reconstruction. (IEEE), pp. 1-4.
19. Wang, J., Liang, J., Cheng, J., Guo, Y., and Zeng, L. (2020). Deep learning based image reconstruction algorithm for limited-angle translational computed tomography. Plos one *15*, e0226963.
20. Hu, D., Liu, J., Lv, T., Zhao, Q., Zhang, Y., Quan, G., Feng, J., Chen, Y., and Luo, L. (2021). Hybrid-Domain Neural Network Processing for Sparse-View CT Reconstruction. IEEE Transactions on Radiation and Plasma Medical Sciences *5*, 88-98. 10.1109/trpms.2020.3011413.
21. Liu, Y., Deng, K., Sun, C., and Yang, H. (2021). A Lightweight Structure Aimed to Utilize Spatial Correlation for Sparse-View CT Reconstruction. arXiv preprint arXiv:2101.07613.
22. Zhang, Q., Hu, Z., Jiang, C., Zheng, H., Ge, Y., and Liang, D. (2020). Artifact removal using a hybrid-domain convolutional neural network for limited-angle computed tomography imaging. Physics in Medicine & Biology *65*, 155010.
23. Wu, W., Hu, D., Niu, C., Yu, H., Vardhanabhuti, V., and Wang, G. (2021). DRONE: Dual-Domain Residual-based Optimization NEtwork for Sparse-View CT Reconstruction. IEEE Trans Med Imaging *40*, 3002-3014. 10.1109/TMI.2021.3078067.
24. Zheng, A., Gao, H., Zhang, L., and Xing, Y. (2020). A dual-domain deep learning-based reconstruction method for fully 3D sparse data helical CT. Physics in Medicine & Biology

*65*, 245030.
25. Cheng, L., Ahn, S., Ross, S.G., Qian, H., and De Man, B. (2017). Accelerated iterative image reconstruction using a deep learning based leapfrogging strategy. pp. 715-720.
26. Chen, H., Zhang, Y., Chen, Y., Zhang, J., Zhang, W., Sun, H., Lv, Y., Liao, P., Zhou, J., and Wang, G. (2018). LEARN: Learned experts' assessment-based reconstruction network for sparse-data CT. IEEE transactions on medical imaging *37*, 1333-1347.
27. Zhang, Y., Chen, H., Xia, W., Chen, Y., Liu, B., Liu, Y., Sun, H., and Zhou, J. (2020). LEARN++: Recurrent Dual-Domain Reconstruction Network for Compressed Sensing CT. arXiv preprint arXiv:2012.06983.
28. Xiang, J., Dong, Y., and Yang, Y. (2021). FISTA-Net: Learning A fast iterative shrinkage thresholding network for inverse problems in imaging. IEEE Transactions on Medical Imaging *40*, 1329-1339.
29. Bello, I., Zoph, B., Vaswani, A., Shlens, J., and Le, Q.V. (2019). Attention augmented convolutional networks. pp. 3286-3295.
30. Cordonnier, J.-B., Loukas, A., and Jaggi, M. (2019). On the relationship between self-attention and convolutional layers. arXiv preprint arXiv:1911.03584.
31. Vaswani, A., Shazeer, N., Parmar, N., Uszkoreit, J., Jones, L., Gomez, A.N., Kaiser, Ł., and Polosukhin, I. (2017). Attention is all you need. pp. 5998-6008.
32. Radford, A., Wu, J., Child, R., Luan, D., Amodei, D., and Sutskever, I. (2019). Language models are unsupervised multitask learners. OpenAI blog *1*, 9.
33. Otter, D.W., Medina, J.R., and Kalita, J.K. (2020). A survey of the usages of deep learning for natural language processing. IEEE Transactions on Neural Networks and Learning Systems *32*, 604-624.
34. Dosovitskiy, A., Beyer, L., Kolesnikov, A., Weissenborn, D., Zhai, X., Unterthiner, T., Dehghani, M., Minderer, M., Heigold, G., and Gelly, S. (2020). An image is worth 16x16 words: Transformers for image recognition at scale. arXiv preprint arXiv:2010.11929.
35. Arnab, A., Dehghani, M., Heigold, G., Sun, C., Lučić, M., and Schmid, C. (2021). Vivit: A video vision transformer. arXiv preprint arXiv:2103.15691.
36. Zhou, H.-Y., Guo, J., Zhang, Y., Yu, L., Wang, L., and Yu, Y. (2021). nnFormer: Interleaved Transformer for Volumetric Segmentation. arXiv preprint arXiv:2109.03201.
37. Liu, Z., Lin, Y., Cao, Y., Hu, H., Wei, Y., Zhang, Z., Lin, S., and Guo, B. (2021). Swin transformer: Hierarchical vision transformer using shifted windows. arXiv preprint arXiv:2103.14030.
38. Ronneberger, O., Fischer, P., and Brox, T. (2015). U-net: Convolutional networks for biomedical image segmentation. (Springer), pp. 234-241.
39. AAPM challenge. Available from: https://www.aapm.org/GrandChallenge/LowDoseCT/.
40. Yu, H., Wang, G., Hsieh, J., Entrikin, D.W., Ellis, S., Liu, B., and Carr, J.J.J.J.o.c.a.t. (2011). Compressive sensing–Based interior tomography: Preliminary clinical application. *35*, 762.
41. Zeng, G. (2015). The fan-beam short-scan FBP algorithm is not exact. Physics in Medicine & Biology *60*, N131.
42. Niu, C., Wang, G., Yan, P., Hahn, J., Lai, Y., Jia, X., Krishna, A., Mueller, K., Badal, A., and Myers, K. (2021). Noise Entangled GAN For Low-Dose CT Simulation. arXiv preprint arXiv:2102.09615.

43. Xu, M., Hu, D., Luo, F., Liu, F., Wang, S., and Wu, W. (2020). Limited-Angle X-Ray CT Reconstruction Using Image Gradient $\ell_0$-Norm With Dictionary Learning. IEEE Transactions on Radiation and Plasma Medical Sciences *5*, 78-87.
44. Wu, W., Zhang, Y., Wang, Q., Liu, F., Chen, P., and Yu, H. (2018). Low-dose spectral CT reconstruction using image gradient $\ell_0$–norm and tensor dictionary. Applied Mathematical Modelling *63*, 538-557.
45. Wu, W., Yu, H., Gong, C., and Liu, F. (2017). Swinging multi-source industrial CT systems for aperiodic dynamic imaging. Optics express *25*, 24215-24235.
46. Yu, H., and Wang, G. (2009). Compressed sensing based interior tomography. Physics in medicine & biology *54*, 2791.
47. Wu, W., Hu, D., An, K., Wang, S., and Luo, F. (2020). A high-quality photon-counting CT technique based on weight adaptive total-variation and image-spectral tensor factorization for small animals imaging. IEEE Transactions on Instrumentation and Measurement *70*, 1-14.
48. Shen, Y., Li, J., Zhu, Z., Cao, W., and Song, Y. (2015). Image reconstruction algorithm from compressed sensing measurements by dictionary learning. Neurocomputing *151*, 1153-1162.
49. Zha, Z., Liu, X., Zhang, X., Chen, Y., Tang, L., Bai, Y., Wang, Q., and Shang, Z. (2018). Compressed sensing image reconstruction via adaptive sparse nonlocal regularization. The Visual Computer *34*, 117-137.
50. Dong, J., Fu, J., and He, Z. (2019). A deep learning reconstruction framework for X-ray computed tomography with incomplete data. PloS one *14*, e0224426.
51. Sobel, I., and Feldman, G. (1968). A 3x3 isotropic gradient operator for image processing. a talk at the Stanford Artificial Project in, 271-272.
52. Liang, T., Jin, Y., Li, Y., and Wang, T. (2020). EDCNN: Edge enhancement-based Densely Connected Network with Compound Loss for Low-Dose CT Denoising. (IEEE), pp. 193-198.
53. Sharifrazi, D., Alizadehsani, R., Roshanzamir, M., Joloudari, J.H., Shoeibi, A., Jafari, M., Hussain, S., Sani, Z.A., Hasanzadeh, F., and Khozeimeh, F. (2021). Fusion of convolution neural network, support vector machine and Sobel filter for accurate detection of COVID-19 patients using X-ray images. Biomedical Signal Processing and Control *68*, 102622.
54. Chen, J., Lu, Y., Yu, Q., Luo, X., Adeli, E., Wang, Y., Lu, L., Yuille, A.L., and Zhou, Y. (2021). Transunet: Transformers make strong encoders for medical image segmentation. arXiv preprint arXiv:2102.04306.
55. Luthra, A., Sulakhe, H., Mittal, T., Iyer, A., and Yadav, S. (2021). Eformer: Edge Enhancement based Transformer for Medical Image Denoising. arXiv preprint arXiv:2109.08044.
56. Liang, J., Cao, J., Sun, G., Zhang, K., Van Gool, L., and Timofte, R. (2021). SwinIR: Image restoration using swin transformer. pp. 1833-1844.

# Supplemental Experimental Procedures

## Appendix A
***More simulated data reconstructions***

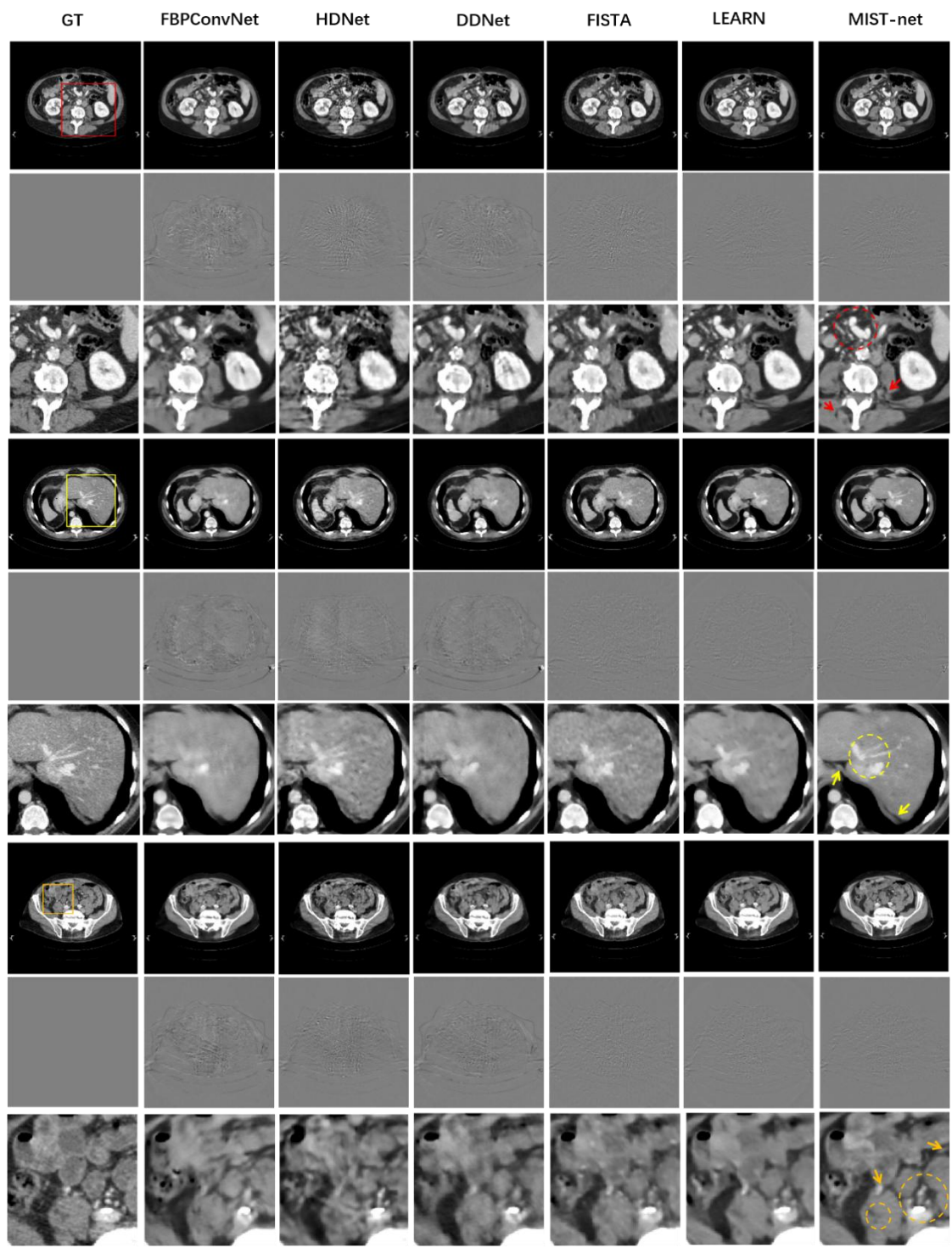

**Figure S1.** More simulated data reconstructions from sparse-view data by using different networks. The 1st-7th columns stand for the FBP reconstruction from full-view data, FBPconvNet, HDNet, DDNet, FISTA, LEARN and MIST-net counterparts from 48 views. The display windows for the reconstructed images are [-160 240] HU.

We also do the experiments on numerical datasets with 64 views. The results can be found in Figure S2.

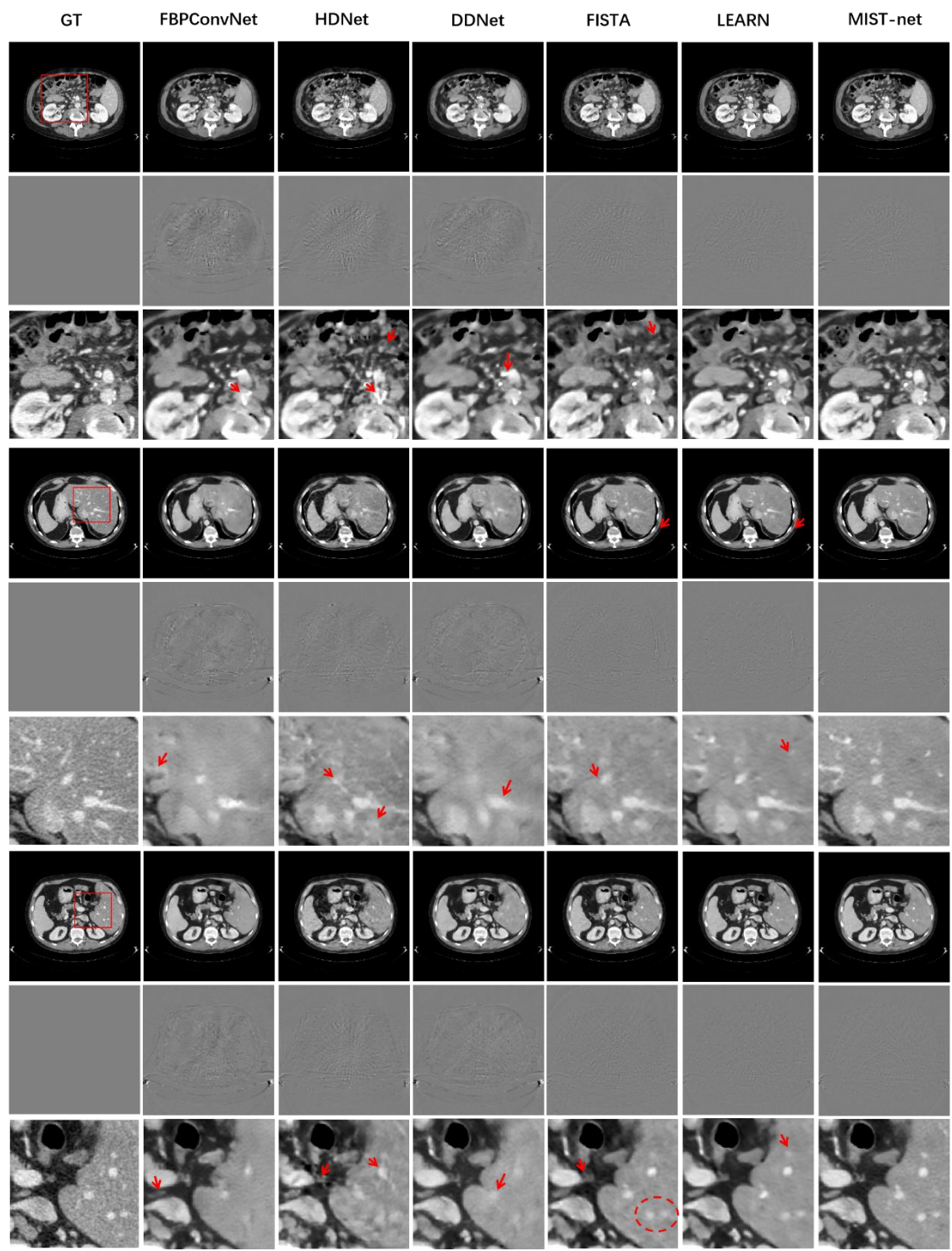

**Figure S2.** The simulated data reconstructions from 64 views by using different networks. The 1st-7th columns stand for the FBP reconstruction from full-view data, FBPconvNet, HDNet, DDNet, FISTA, LEARN and MIST-net. The display windows for the reconstructed images are [-160 240] HU.

# Appendix B
*The statistical quantitative evaluations*

The statistical quantitative evaluations result from testing datasets were computed in terms of RMSE, PSNR and SSIM, and their results were summarized in Table S1. It can be seen that our MIST-net can obtain the best quantitative statistical results in terms of mean and standard deviation than other competitors.

Table S1

Quantitative evaluation of 64 projections reconstruction results from simulated testing datasets

| Views | Methods | RMSE | PSNR | SSIM |
|---|---|---|---|---|
| 64 | FBPconvNet | 25.9083 ± 4.5941 | 38.6168 ± 1.6026 | 0.9634 ± 0.0174 |
|  | HDNet | 22.8740 ± 4.2488 | 39.2675 ± 1.2331 | 0.9670 ± 0.0144 |
|  | DDNet | 24.1746 ± 4.7592 | 39.7018 ± 1.3685 | 0.9655 ± 0.0163 |
|  | FISTA | 17.4884 ± 3.1788 | 42.0121 ± 1.2369 | 0.9769 ± 0.0100 |
|  | LEARN | 15.9171 ± 3.6576 | 42.8622 ± 1.3997 | 0.9814 ± 0.0123 |
|  | MIST-net | **14.5492 ± 3.2807** | **43.6418 ± 1.4070** | **0.9844 ± 0.00101** |

In addition to the performance metrics, we also compare the complexity and runtime of all competitors. The test time represents the time taken to predict a total of 391 test datasets. Table S2 shows that our MIST-net train and test much faster than LEARN and FISTA.

Table S2

Time-consuming comparison of different methods (48 projections)

| Views |  | FBPconvNet | HDNet | DDNet | FISTA | LEARN | MIST-net |
|---|---|---|---|---|---|---|---|
| 48 | Epochs | 150 | 150(×2) | 100 | 40 | 40 | 40 |
|  | Input Size | 512×512 | 48×880 | 512×512 | 48×880 | 48×880 | 48×880 |
|  | Output Size | 512×512 | 512×512 | 512×512 | 512×512 | 512×512 | 512×512 |
|  | #Param. | 9.8M | 9.8M(×2) | 6.4M | 1.3M | 3.0M | 12.0M |
|  | Train Time | 21.3h | 37.5h | 41.2h | 30.1h | 83.1h | 38.7h |
|  | Test Time | 40s | 63s | 66s | 105s | 316s | 98s |

# Appendix C
***More noise experiment results***

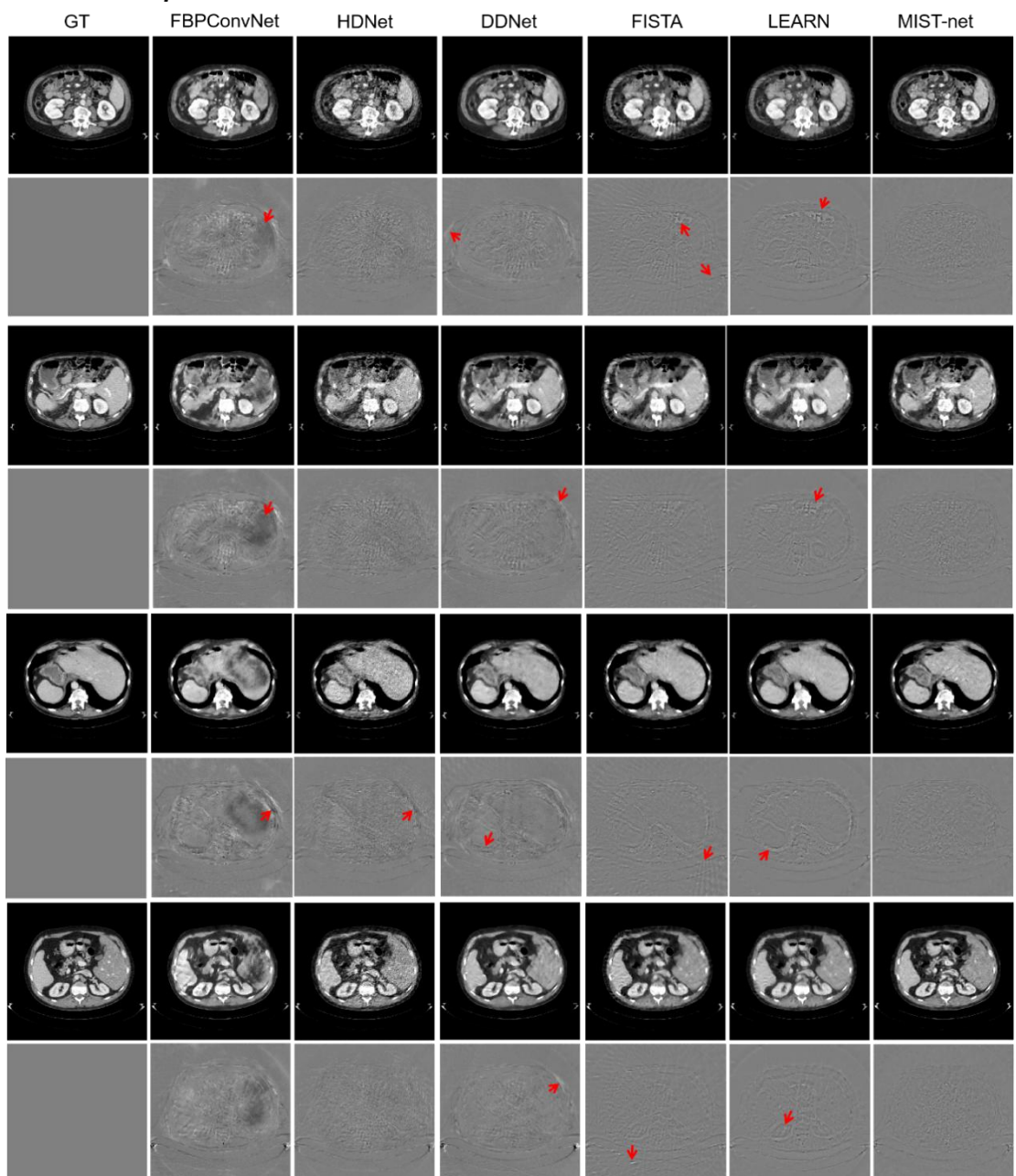

**Figure S3.** The generalization of different deep reconstruction networks against noise on simulation datasets. The 1st-7th columns stand for the ground truth, FBPconvNet, HDNet, DDNet, FISTA, LEARN and MIST-net from 48 views. The display window of reconstructed images is [-160 240] HU.

# Appendix D
***More network details***

Table S3

Parametric structure for all the layers in the encoder-decoder block

| Layers | Parameters | Input Channel | Output Channel |
|---|---|---|---|
| A0 | 3×3Conv+BN+ReLU | 1 | 32 |
|  | 3×3Conv+BN+ReLU |  |  |
| A1 | 2×2 MaxPooling | 32 | 64 |
|  | 3×3Conv+BN+ReLU |  |  |
|  | 3×3Conv+BN+ReLU |  |  |
| A2 | 2×2 MaxPooling | 64 | 128 |
|  | 3×3Conv+BN+ReLU |  |  |
|  | 3×3Conv+BN+ReLU |  |  |
| A3 | 2×2 MaxPooling | 128 | 256 |
|  | 3×3Conv+BN+ReLU |  |  |
|  | 3×3Conv+BN+ReLU |  |  |
| A4 | 2×2 MaxPooling | 256 | 512 |
|  | 3×3Conv+BN+ReLU |  |  |
|  | 3×3Conv+BN+ReLU |  |  |
| B0 | Upsample+3×3Conv+BN+ReLU | 512 | 256 |
|  | Concatenation |  |  |
|  | 3×3Conv+BN+ReLU |  |  |
|  | 3×3Conv+BN+ReLU |  |  |
| B1 | Upsample+3×3Conv+BN+ReLU | 256 | 128 |
|  | Concatenation |  |  |
|  | 3×3Conv+BN+ReLU |  |  |
|  | 3×3Conv+BN+ReLU |  |  |
| B2 | Upsample+3×3Conv+BN+ReLU | 128 | 64 |
|  | Concatenation |  |  |
|  | 3×3Conv+BN+ReLU |  |  |
|  | 3×3Conv+BN+ReLU |  |  |
| B3 | Upsample+3×3Conv+BN+ReLU | 64 | 32 |
|  | Concatenation |  |  |
|  | 3×3Conv+BN+ReLU |  |  |
|  | 3×3Conv+BN+ReLU |  |  |
| B4 | $1 \times 1$ Convolution | 32 | 1 |
| B5 | Residual Connection | 1 | 1 |

Table S4

Parametric structure for all the layers in the edge enhancement Rec-network

| Layers | Parameters | Input Channel | Output Channel |
|---|---|---|---|
| C0 | Sobel Convolution | 1 | 32 |
|  | Concatenation |  | 33 |
| C1 | 3×3Conv+ReLU | 33 | 32 |
|  | 3×3Conv+ReLU | 32 | 32 |
|  | Concatenation |  | 65 |
| C2 | 3×3Conv+ReLU | 65 | 32 |
|  | 3×3Conv+ReLU | 32 | 32 |
|  | Concatenation |  | 65 |
| C3 | 3×3Conv+ReLU | 65 | 32 |
|  | 3×3Conv+ReLU | 32 | 32 |
|  | Concatenation |  | 65 |
| C4 | 3×3Conv+ReLU | 65 | 32 |
|  | 3×3Conv+ReLU | 32 | 32 |
|  | Concatenation |  | 65 |
| C5 | 3×3Conv+ReLU | 65 | 32 |
|  | 3×3Conv+ReLU | 32 | 32 |
|  | Concatenation |  | 65 |
| C6 | 3×3Conv+ReLU | 65 | 32 |
|  | 3×3Conv+ReLU | 32 | 32 |
|  | Concatenation |  | 65 |
| C7 | 3×3Conv+ReLU | 65 | 32 |
|  | 3×3Conv+ReLU | 32 | 32 |
|  | Concatenation |  | 65 |
| C8 | 3×3Conv+ReLU | 65 | 32 |
|  | 3×3Conv | 32 | 1 |
|  | Residual Connection | 1 | 1 |

Table S5

Parametric structure for all the layers in the Swin Rec-former

| Layers | Parameters | Window Size | Head Numbers | Head Numbers |
|---|---|---|---|---|
| D0 | 3×3convolution | | | |
| D1 | Patch-Embed Block | 8×8 | 6 | 1×1 |
| | Swin Transformer Block (×6) | | | |
| | Patch-UnEmbed Block | | | |
| | 3×3convolution+ Residual Connection | | | |
| D2 | Patch-Embed Block | 8×8 | 6 | 1×1 |
| | Swin Transformer Block (×6) | | | |
| | Patch-UnEmbed Block | | | |
| | 3×3convolution+ Residual Connection | | | |
| D3 | Patch-Embed Block | 8×8 | 6 | 1×1 |
| | Swin Transformer Block (×6) | | | |
| | Patch-UnEmbed Block | | | |
| | 3×3convolution+ Residual Connection | | | |
| D4 | Patch-Embed Block | 8×8 | 6 | 1×1 |
| | Swin Transformer Block (×6) | | | |
| | Patch-UnEmbed Block | | | |
| | 3×3convolution+ Residual Connection | | | |
| D5 | 3×3convolution+ Residual Connection | | | |